# MedYOLO: A Medical Image Object Detection Framework

## Authors

Joseph Sobek [a], Jose R. Medina Inojosa M.D. M.Sc. [b,c], Betsy J. Medina Inojosa M.D. [b], S. M. Rassoulinejad-Mousavi [a], Gian Marco Conte [a], Francisco Lopez-Jimenez M.D., M.Sc., M.B.A. [b], Bradley J. Erickson [a]

Affiliations:

[a] Department of Radiology, Mayo Clinic, Rochester, MN.

[b] Department of Cardiovascular Medicine, Mayo Clinic, Rochester, MN.

[c] Division of Epidemiology, Department of Quantitative Health Sciences, Mayo Clinic, Rochester, MN.

## Abstract

Artificial intelligence-enhanced identification of organs, lesions, and other structures in medical imaging is typically done using convolutional neural networks (CNNs) designed to make voxel-accurate segmentations of the region of interest. However, the labels required to train these CNNs are time-consuming to generate and require attention from subject matter experts to ensure quality. For tasks where voxel-level precision is not required, object detection models offer a viable alternative that can reduce annotation effort. Despite this potential application, there are few options for general purpose object detection frameworks available for 3-D medical imaging. We report on MedYOLO, a 3-D object detection framework using the one-shot detection method of the YOLO family of models and designed for use with medical imaging. We tested this model on four different datasets: BRaTS, LIDC, an abdominal organ Computed Tomography (CT) dataset, and an ECG-gated heart CT dataset. We found our models achieve high performance on commonly present medium and large-sized structures such as the heart, liver, and pancreas even without hyperparameter tuning. However, the models struggle with very small or rarely present structures.

Keywords: Object detection, medical imaging, computed tomography, magnetic resonance, convolutional neural network, deep learning

## Background

The standard approach for object localization in 3-D medical imaging uses segmentation models to create voxel-by-voxel annotations for the objects of interest. While this approach enables models to have great accuracy, it has several downsides. Generating voxel-accurate annotations for medical imaging is a time-consuming process that often requires multiple experts to verify label quality. Due to variability between annotators, medically accurate segmentations of organs or lesions can have issues with indefinite structure boundaries, potentially including irrelevant or excluding relevant information in nearby tissue. Even with high-quality labels, segmentation models can struggle to accurately label target structure boundaries and often require post-processing to fill missing internal volumes and remove spurious predicted objects. Altogether this makes segmentation models cost prohibitive to train while potentially limiting the predictive power of downstream diagnostic or classification models.

Object detection models offer an alternative that avoids these issues for tasks where voxel-specific accuracy is not required. Despite this, relatively few options are available for object detection in 3-D medical imaging. 2-D object detection models intended for use with photographs, such as YOLO [1], can provide bounding boxes with slice-by-slice accuracy. However, 2-D models require cumbersome conversion processes for both their input, to resolve the incompatibility of 3-D input data with the model, and their output, to reconstruct the predictions for use in downstream, 3-D tasks. This conversion process also discards 3-D spatial information, which may be helpful when detecting complex structures.

We have developed a 3-D object detection framework for medical imaging to mitigate this sparsity of options. MedYOLO is based on the Ultralytics YOLOv5 [2] detection model and offers high accuracy for medium and large-sized structures. The framework has native compatibility with NIfTI imaging. Our results show its single-shot approach works particularly well on structures that patch-based or sliding window approaches, such as that used by nnDetection [3], appear to struggle with.

## Methods

### Data

We tested MedYOLO on four types of medical imaging. These included: 1) FLAIR scans from the BRaTS 2021 Task 1 dataset (1000 training scans, 251 validation) [4, 5, 6, 7, 8] using the whole tumor segmentation mask to generate bounding boxes. 2) LIDC lung nodule dataset (689 training scans, 173 validation) [9, 10, 11] using two different sets of labels, one placing bounding boxes around individual nodules and the other using a single bounding box containing every nodule in each scan. These were used to test the model's detection of very small objects and diffuse structures respectively. 3) a proprietary abdominal organ segmentation dataset [12] (60 training scans, 15 validation). From this, bounding boxes were generated for the following organs: left kidney, right kidney, spleen, pancreas, diaphragm, bladder, uterus, prostate, aorta, spinal cord, stomach, and liver. For this dataset, an additional preprocessing step was used to generate rotated training examples. Training scans were rotated in the axial view using five different base rotation angles for each example (0, ±8, and ±17 degrees) with an additional random jitter of ±3 degrees. An identical rotation was applied to each image's segmentation mask, then bounding box labels for the rotated example were generated. The original, unrotated examples were also included in the training set to mitigate any spurious signals from the interpolation algorithm used for the rotation. In total, this resulted in 360 training examples for the abdominal dataset. 4) a clinically indicated and randomly selected ECG-gated cardiac Computed Tomography (CT) dataset (648 training scans, 163 validation) curated with data from Mayo Clinic and annotated by the authors using the RILContour application [13] for which we attempted to predict a bounding box around the heart and thoracic aorta. Each of these datasets was split at the patient level to prevent data leakage between the training and validation sets.

**Network Design**

MedYOLO is modeled after the YOLOv5 framework but rewritten for use with 3-D medical imaging. It accepts NIfTI files as input and includes normalization functions for CT and Magnetic Resonance (MR) scans. Like YOLOv5, it is an anchor-based detection model created from a convolutional neural network (CNN) with a terminal detection layer that connects to intermediary layers earlier in the network. The primary difference within the CNN, compared to YOLOv5, is the replacement of 2-D neural network layers with their 3-D versions. The width and number of MedYOLO's layers are configurable using yaml files, with small, medium, and large versions included in the repository. Several of YOLOv5's key library dependencies, such as OpenCV, are incompatible with 3-D data, and removing these dependencies required significant changes to the data handling pipeline and the removal of some augmentation routines. A limited portion of the YOLOv5 codebase is compatible with the additional dimension, and this code was retained to improve the interpretability of the MedYOLO framework.

As in earlier versions of YOLO [1], MedYOLO calculates priors for its anchor boxes using k-means clustering on the training set labels. Our initial tests using the elbow method determined six anchor boxes, in contrast to the three used by YOLOv5, to be the appropriate number. This was left fixed for every dataset tested but is a configurable hyperparameter.

The MedYOLO CNN requires cubic inputs analogous to the square inputs used by 2-D YOLO models. Since medical imaging is typically anisotropic, we use trilinear interpolation to reshape input data into cubes of a user-configurable size. To balance batch size against available GPU resources and avoid errors caused by undersized inputs, a side length of 350 pixels per side was chosen for most of our tests. This gives us a final feature map size of 11x11x11 at the bottom of the CNN. Cubes with side length 512 were also tested for the LIDC dataset, which gave a final feature map size of 16x16x16. See Table 1 for the corresponding GPU footprints.

| Model | Batch Size | Cube Length (voxels) | VRAM (GB) |
| --- | --- | --- | --- |
| MedYOLO-S | 8 | 350 | 35 |
| MedYOLO-S | 2 | 512 | 26 |
| MedYOLO-L | 2 | 350 | 25 |

Table 1: VRAM consumption for different model sizes at different input data scales.

An example NIfTI scan, which for the tested datasets typically has a shape between 512x512x40 and 512x512x100, loaded into the pipeline is first converted into a PyTorch Tensor. This Tensor is then transposed from (X, Y, Z) to (Z, X, Y) such that it has a shape 40x512x512. The Tensor is then interpolated into a cube with a shape of 350x350x350. Augmentation, as detailed in the next section, is then applied. Finally, a modality appropriate normalization is applied to the Tensor. The framework allows users to apply their own normalization and preprocessing routines if desired.

**Training**

The MedYOLO training process largely matches that of YOLOv5. We use nearly identical hyperparameters to those used by YOLOv5 for training from scratch, the only exception being hyperparameters to configure the added augmentation routines. Several of YOLOv5's augmentation routines needed to be removed, such as color value shifts and random perspective shifts, as they are inapplicable to grayscale data or rely on 2-D imaging libraries. Three forms of live augmentation were used: random cutout augmentation, which replaces blocks of random size within the image with random noise, random translation augmentation, and random zoom augmentation. We trained the small version of MedYOLO on each dataset for 1000 epochs, with early stopping set to occur after 200 epochs passed with no improvement.

MedYOLO uses a version of YOLOv5's composite loss function adapted for 3-D volumes. The bounding box loss component compares the intersection over union (IoU) and the distance between the centers of the predicted and target bounding boxes. The objectness component, which trains the model to evaluate its predictions by comparing their IoUs to the model's confidence level, is calculated using binary cross-entropy. Finally, the classification loss uses binary cross-entropy on the predicted class. The repository includes focal loss options for the latter two components, but these were not used for the results reported in this paper.

The nnDetection framework [3] was trained on the same datasets with identical training and test splits for comparison purposes. Like MedYOLO, nnDetection is a general-purpose medical imaging object detection framework intended for use with 3-D NIfTI images, however it uses a sliding window object detection method as opposed to MedYOLO's single-shot approach. The nnDetection framework is also self-configuring, featuring automatic preprocessing, augmentation, and 5-fold cross-validation. Because nnDetection's cross-validation is automatically configured during preprocessing and the framework does not track patient IDs, we did not use the rotated examples for the abdominal CT dataset with nnDetection to avoid having examples from a single patient spread between multiple folds.

## Results

In Table 2, we report the best validation metrics after training both MedYOLO and nnDetection. Where not otherwise specified, the small version of MedYOLO (MedYOLO-S) is reported on. For the BRaTS dataset early stopping occurred at epoch 675, for the ECG gated cardiac dataset early stopping occurred at epoch 938, and for the abdominal dataset the small model trained for 1000 epochs. We also trained the large version of MedYOLO (MedYOLO-L) on the abdominal dataset, for which early stopping occurred at epoch 968, and on BRaTS, on which it failed to converge. We report the mean average precision (mAP) at IoU 0.5 for both frameworks. For MedYOLO we report a 3-D version of the COCO metric [14], which calculates the average mAP from IoUs ranging between 0.5 to 0.95 with an interval of 0.05, as reported by YOLOv5. The nnDetection framework only reports mAP in IoU intervals of 0.1 up to an IoU of 0.9, so for it we report a slightly more relaxed metric averaging mAP for IoUs between 0.5 and 0.9 with an interval of 0.1.

|  | MedYOLO mAP@0.5 | MedYOLO mAP@0.5:0.95 | nnDetection mAP@0.5 | nnDetection mAP@0.5:0.9 |
|---|---|---|---|---|
| BRaTS | 0.861 | 0.431 | 0.836 | 0.488 |
| Gated Cardiac | 0.995 | 0.852 | 0 | 0 |
| Abdominal | .683 | .203 | .208 | .064 |
| Abdominal (MedYOLO-L) | .715 | .246 | .208 | .064 |

Table 2: mAP comparisons between MedYOLO and nnDetection. For a further breakdown of the classes in the Abdominal dataset, see Table 1S in the supplementary figures.

Our results show MedYOLO has particularly good performance on large, common structures, exceeding nnDetection's performance in most cases. The Gated Cardiac dataset in particular shows significantly different behavior between the frameworks. While MedYOLO nearly completely captures the target volume, nnDetection shows poor results at high IoUs. Further investigation showed predictions from the nnDetection framework are highly localized to the target volume but span little of the target volume. This creates many predictions with IoUs between 0.1 and 0.4, but for this dataset none with IoUs above 0.5.

Despite good results for mid-size and larger structures, MedYOLO struggles to identify very small structures and uncommon classes. During training, MedYOLO failed to demonstrate any learning progress for the LIDC dataset in any tested configuration. We attribute this to the coarse final feature maps of MedYOLO's CNN limiting the model's capacity to detect very small structures.

Table 1S shows a more detailed breakdown of MedYOLO's performance on the Abdominal dataset compared to nnDetection. While MedYOLO generally outperformed nnDetection in the abdomen, the latter performed significantly better at identifying the uterus, a class present in only 17 out of 60 abdominal training examples and 4 out of 15 validation examples. The large MedYOLO model, MedYOLO-L, provides general improvements to performance on the Abdominal dataset at the cost of slower training. However, its performance on the uterus is still substantially worse than nnDetection's performance. As discussed in the training strategy, MedYOLO-L also failed to converge on the BRaTS dataset despite similar performance by MedYOLO-S to that of nnDetection.

## Discussion

MedYOLO shows promising results for use as an object detector for 3-D medical imaging, capable of achieving good results using a relatively simple, one-shot detection method. Compared to nnDetection, MedYOLO shows similar performance on medium-sized structures and much better performance on large-sized structures than the sliding window method used by nnDetection. However, in our analyses it failed to detect very small or diffuse structures. Achieving high mAP on a variety of structures without hyperparameter tuning suggests MedYOLO has some robustness against suboptimal hyperparameter choices. The framework's ability to quickly and accurately find medium and large-sized structures makes it ideal for use in machine learning pipelines to localize data to relevant volumes before it is passed to downstream models.

Despite the impressive performance for some tasks, the current implementation has significant potential for improvement. One of the biggest avenues for improvement would be additional augmentation routines. In particular, the performance on BRaTS tumor detection, with MedYOLO-S finishing training early despite comparably modest results and MedYOLO-L failing to converge, suggests the model may be missing pertinent information. Augmentation could extend the training process to achieve higher performance on this and other datasets. The obstacle we encounter is a relative lack of efficient and easily implementable augmentation routines for 3-D data, causing many standard live augmentation strategies to slow down training speeds significantly. Pre-generating augmented training data avoids this issue, but this increases storage requirements drastically due to the large memory footprint of medical imaging.

Another modification that may improve performance is switching the algorithm we use for reshaping data. Trilinear interpolation allows us to smoothly transform 3-D input data into a cubic shape but does not improve the information available in the input. A more sophisticated resampling method, such as super-resolution, could provide extra detail and add value to the slices created during the reshaping process.

The requirement to reshape input data into cubic volumes is probably the biggest weakness of our pipeline. Increasing the number of input slices, often by an order of magnitude, dramatically increases the computational resources demanded by our model. In addition, balancing batch memory constraints against the accuracy of batch statistics requires us to use smaller cubes and reduce the axial resolution. Gradient accumulation over several batches could help resolve this. We did not experiment with this because we achieved adequate performance without it on most of the datasets tested. While it may have been useful for the LIDC dataset, the lack of any training progress even when giving the model the full axial resolution of the images suggests this is unlikely. A related issue comes from medical imaging datasets in general, which often consist of images with a variable number of slices. Reshaping these datasets into a fixed, cubic size risks distorting input images in unpredictable ways relative to one another. During inference, scans with unusually few slices are what the model primarily fails to accurately predict bounding boxes for, even for the ECG-gated cardiac dataset on which our validation metrics are extremely high.

Early during development, substantial effort was applied to remove the cubic shape requirement from the MedYOLO framework, at least for the added depth dimension, but this is necessary for the following reasons. Firstly, the depth and number of pooling layers in the MedYOLO and YOLO architectures impose minimum resolution requirements on every dimension. The resolution chosen for most of the discussed datasets, 350x350x350, is close to this minimum resolution. The more difficult to resolve problem is that YOLOv5 models require square input data to simplify anchor-based prediction. This becomes a requirement for cubic inputs when transforming these 2-D models into 3-D. To handle rectangular inputs, YOLOv5 resamples and zero-pads input into square shapes which reduces the distortion that occurs during resampling. This process is computationally cheap for 2-D data, but the larger memory footprint and highly anisotropic shapes of medical imaging make it impractical to zero-pad 3-D input data into cubes, limiting us to resampling techniques.

Despite these limitations, MedYOLO demonstrates that one-shot, anchor-based approaches can achieve high performance on 3-D medical object detection tasks. However, future frameworks may perform better using this YOLO-like approach in a 2.5-D paradigm. This would enable new frameworks to maintain the native resolution of their input data without compromising batch size or introducing distortion from reshaping. The primary downside of a 2.5-D approach compared to a 3-D approach is that additional annotation effort is required to maintain bounding box accuracy over large structures, but they still ease annotation effort relative to voxel-accurate segmentation.

## Conclusion

We report the development of MedYOLO, an anchor-based object detection framework written for use with 3-D medical imaging. MedYOLO matches or exceeds nnDetection's performance at locating medium to large-sized structures even without hyperparameter optimization. Areas for future improvement to the pipeline include adding augmentation routines as efficient 3-D algorithms are developed and incorporating more sophisticated interpolation methods for resampling input data into the required cubic shapes.

## Acknowledgments

The authors acknowledge the National Cancer Institute and the Foundation for the National Institutes of Health, and their critical role in the creation of the free publicly available LIDC/IDRI Database used in this study.

This preprint has not undergone peer review or any post-submission improvements or corrections. The Version of Record of this article is published in Journal of Imaging Informatics in Medicine, and is available online at https://doi.org/10.1007/s10278-024-01138-2

## Declarations

Code Availability: Code for MedYOLO is available at: https://github.com/JDSobek/MedYOLO

Data Availability: BRaTS and LIDC data are available from TCIA. To protect patient privacy, the abdominal and cardiac datasets are not available to share.

Funding: This work was supported by Mayo Clinic.

Ethics Approval: This retrospective study was reviewed and deemed exempt by Mayo Clinic's Institutional Review Board.

Consent to Participate: Informed consent was waived by ethics groups.

Consent to Publication: Informed consent was waived by ethics groups.

Conflict of Interest: The authors declare no competing interests.

## Supplemental Figures

| MO test set performance per class | nnDetection | | MedYOLO-S | | MedYOLO-L | |
|---|---|---|---|---|---|---|
| | mAP@0.5 | mAP@0.5:0.9 | mAP@0.5 | mAP@0.5:0.95 | mAP@0.5 | mAP@0.5:0.95 |
| Left Kidney | .308 | .118 | .49 | .115 | .866 | .254 |
| Right Kidney | .251 | .061 | .781 | .199 | .869 | .394 |
| Spleen | .124 | .034 | .836 | .198 | .698 | .158 |
| Pancreas | .090 | .025 | .868 | .202 | .85 | .225 |
| Diaphragm | 0 | 0 | .752 | .338 | .82 | .431 |
| Bladder | .725 | .223 | .647 | .247 | .702 | .245 |
| Uterus | .629 | .202 | .138 | .019 | .201 | .073 |
| Prostate | .277 | .081 | .75 | .228 | .897 | .219 |
| Aorta | .01 | .004 | .634 | .12 | .593 | .166 |
| Spinal Cord | 0 | 0 | .567 | .14 | .423 | .159 |
| Stomach | .085 | .026 | .791 | .2 | .724 | .197 |
| Liver | 0 | 0 | .937 | .435 | .942 | .431 |

Table 1S: Breakdown of the abdominal dataset mAPs by class for each model.

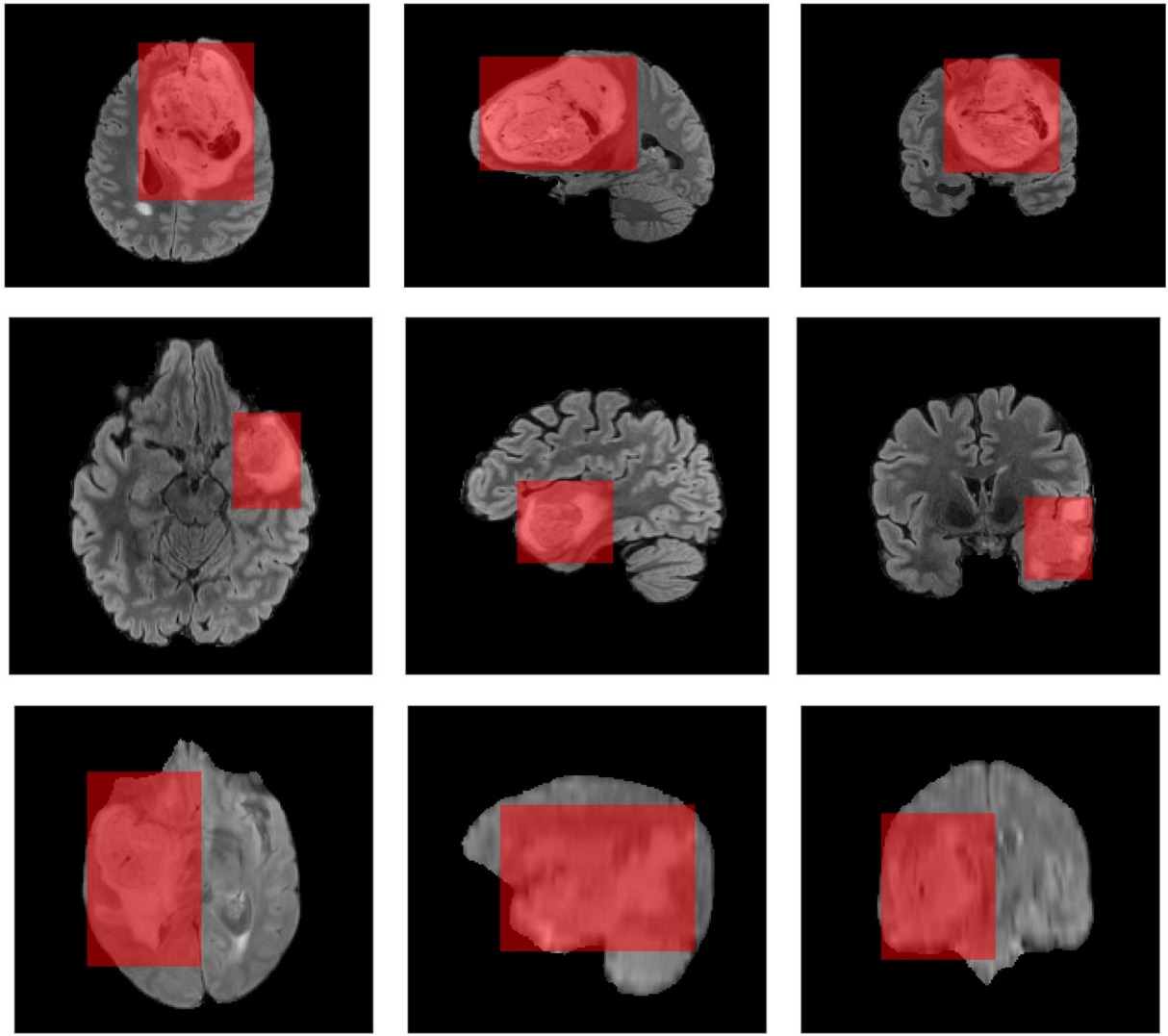

Figure 1S: Bounding box segmentations for BRaTS validation data on three different scans.

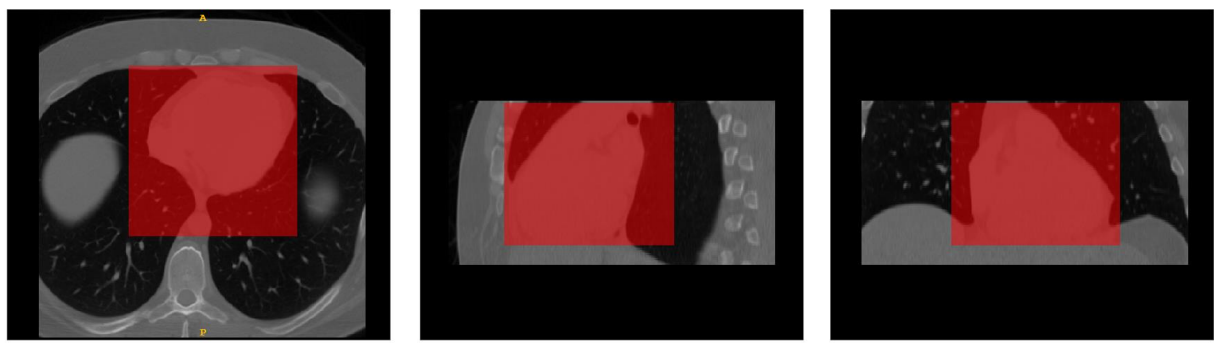

Figure 2S: Bounding box segmentations for ECG gated cardiovascular measurement.

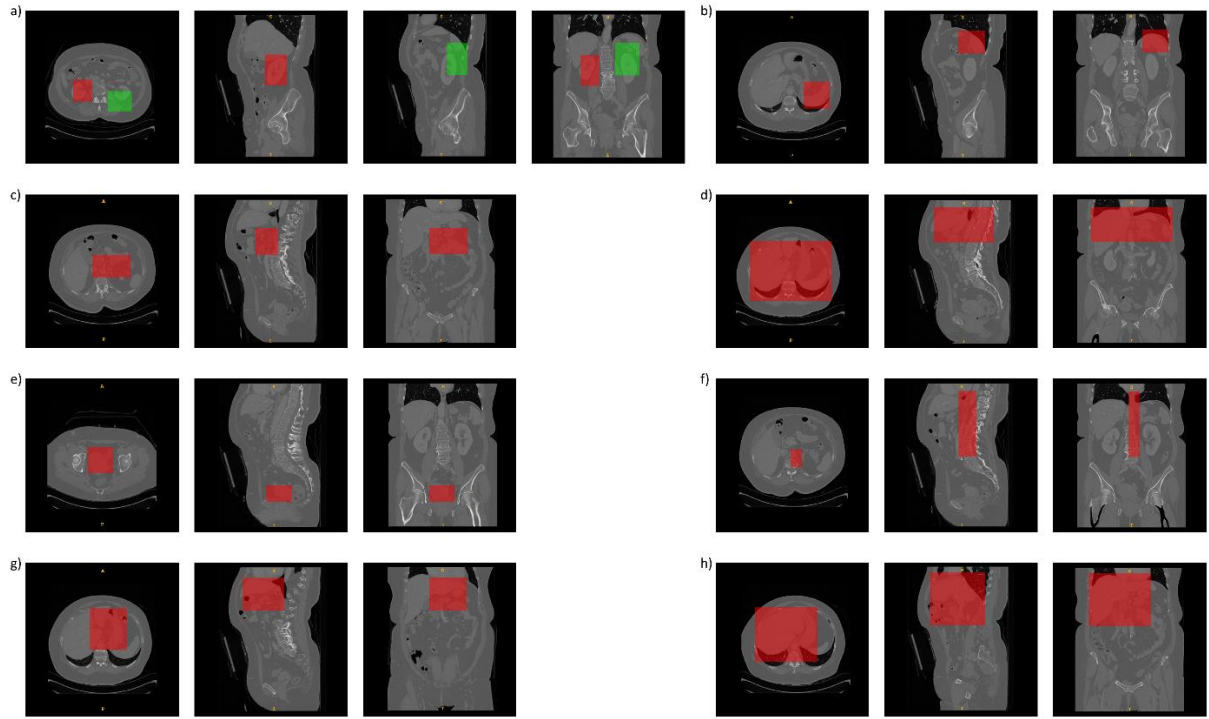

Figure 3S: Bounding box segmentations for several organs in the abdominal dataset. a) Left and Right Kidneys b) Spleen c) Pancreas d) Diaphragm e) Bladder f) Aorta g) Stomach h) Liver.

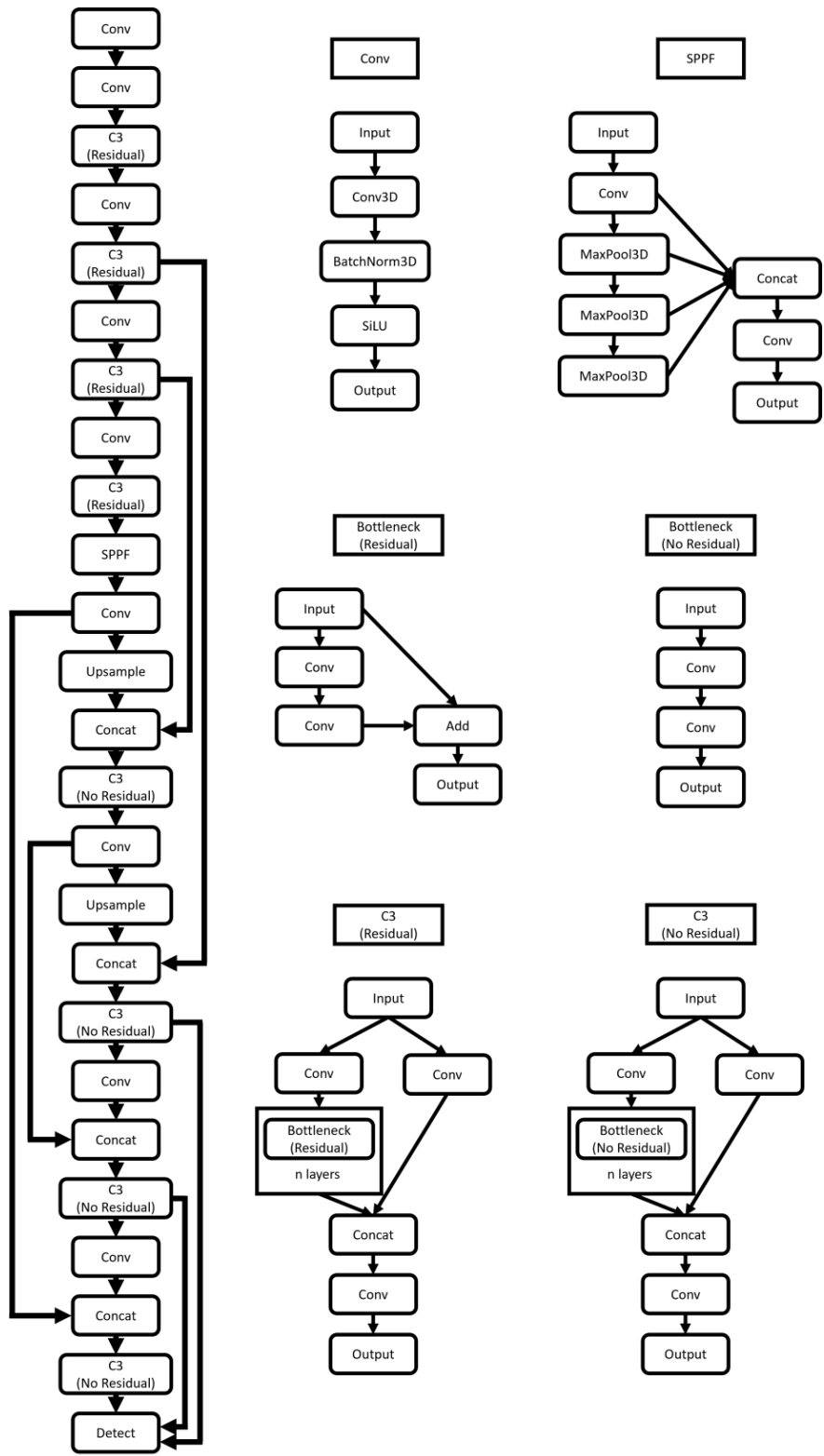

Figure 4S: Architectural diagram of MedYOLO and its component modules. Due to its complexity the Detect module is easier understood through code and not displayed here.